\documentclass{article}
\usepackage{graphicx,amsmath,amsfonts,theorem}
\newtheorem{definition}{Definition}
\newtheorem{proposition}{Proposition}
\newtheorem{corollary}{Corollary}
\newtheorem{lemma}{Lemma}
\begin{document}

%
%

\newcommand{\R}{\mathbb R}
\newcommand{\N}{\mathbb N}
\newcommand{\eps}{\varepsilon}
\newcommand{\To}{\longrightarrow}
\newcommand{\n}{\underline{n}}

\newcommand{\gap}{\mathrm{gap}}
\newcommand{\range}{\mathrm{range}}
\newcommand{\sign}{\mathrm{sign}}
\newcommand{\modd}{\mathrm{mod}}
\newcommand{\epslow}{\eps_\mathrm{low}}
\newcommand{\epshigh}{\eps_\mathrm{high}}

\title{About the Power to \\ Enforce and Prevent Consensus \\ by Manipulating Communication Rules}
\author{Jan Lorenz\\Department of Mathematics and Computer Science\\ Universit\"at Bremen, Bibliothekstra{\ss}e\\
 28359 Bremen, Germany, math@janlo.de \vspace{.2cm} \\
Diemo Urbig\\School of Business and Economics and \\  Department of Computer
Science \\Humboldt-Universit\"at zu Berlin, Unter den Linden 6, \\ 10099
Berlin, Germany, 
diemo@urbig.org}
\date{Preprint for Advances in Complex Systems, \\Vol. 10, No. 2 (2007) 251-269}

\maketitle


\begin{abstract}
We explore the possibilities of enforcing and preventing consensus
in continuous opinion dynamics that result from modifications in the communication
rules. We refer to the model of Weisbuch and Deffuant, where $n$
agents adjust their continuous opinions as a result of random
pairwise encounters whenever their opinions differ not more than a
given bound of confidence $\eps$. A high $\eps$ leads to consensus,
while a lower $\eps$ leads to a fragmentation into several opinion
clusters.
We drop the random encounter assumption and ask: How small may
$\eps$ be such that consensus is still possible with a certain
communication plan for the entire group? Mathematical analysis
shows that $\eps$ may be significantly smaller than in the random
pairwise case. On the other hand we ask: How large may $\eps$ be
such that preventing consensus is still possible? In answering
this question we prove Fortunato's simulation result that
consensus cannot be prevented for $\eps>0.5$ for large groups.
Next we consider opinion dynamics under different individual 
strategies and examine their power to increase the chances
of consensus. One result is that balancing agents increase chances
of consensus, especially if the agents are cautious in adapting
their opinions. However, curious agents increase chances of consensus
only if those agents are not cautious in adapting their opinions.
\end{abstract}


\section{Introduction}

What happens if people meet and discuss their opinions regarding a
political party, a brand, or a new product? Generally, when people meet they
influence one another and as a consequence may change their opinions. 
Such opinion formation processes are at the heart of models that explain voting
behavior as well as models of innovation diffusion (see for instance \cite{Deffuant2001}).

If people who are assumed to have opinions toward something meet and discuss, 
they may adapt their opinions towards the other agent's opinion and reach a compromise or they may move away from consensus when their initial positions are too different as well as they could ignore each other. For simplification we only consider one-dimensional opinions such that they can be represented by real numbers between zero and one. We will only examine compromising agents under bounded confidence, which implies that individuals who differ too much in their opinions do not affect and thus ignore each other. This assumption mirrors the psychological concept of selective exposure, where people tend to perceive their environment in favor of their own opinions and thereby avoid communication with people with conflicting opinions. However, if agents do not ignore each other, then they get closer in their opinions. 
Such systems of agents, who update their opinions via averaging with other sufficiently similar opinions, are referred to as systems of \emph{continuous opinion dynamics under bounded confidence}.
Models following this paradigm were proposed by Hegselmann and Krause \cite{Krause2000,
Hegselmann2002} and Weisbuch, Deffuant, and others
\cite{Deffuant2000, Weisbuch2002}. In the Hegselmann and Krause model (HK model)
every agent perceives the opinions of every other agent and builds
his new opinion as an average of sufficiently close opinions. Thereby Hegselmann and Krause added the
assumption of bounded confidence to a previous linear opinion dynamics model
by DeGroot \cite{DeGroot1974, Lehrer1981}. Hegselmann and Krause's main question was
what conditions related to the bounded confidence, in other words the degree of open mindedness, are necessary for a consensus to be reached.

While in Hegselman and Krause all agents interact simultaneously, 
the agents in the model by Weisbuch and Deffuant (WD model) 
engage in random pairwise encounters. Several other extensions 
(e.g.\ in \cite{Hegselmann2004}) and a combination of both models, the HK model and the WD model, 
\cite{Urbig2004} have been analyzed. A
model which includes the centrifugal forces of rejecting agents is proposed 
by Jager \cite{Jager2005}. Opinion dynamics models were also examined in 
incompletely linked networks, for instance in scale-free networks
\cite{Fortunato2004a, Amblard2004}. 

While the conditions necessary for consensus were already 
examined regarding the bounded confidence, we will explore 
conditions affecting the rules of communication in 
the sense of who talks with whom. Even if we regard a completely 
linked society as given and thus look at the WD model, this
model has an unexplored free parameter in the
order of who communicates with whom at what time. We will call
rules that modify this order the \emph{communication regime}. 
Studying this parameter is the aim of this paper. 
Considering the complexity of human organizations and the different institutions that foster or manipulate the communication regime, makes immediately clear why this question is of relevance. We will see that although the bounds of confidence have a significant impact also the factors that control the communication regime significantly affect the emergence of consensus or dissent.
Thus, our two leading questions are: To what extent does the
communication regime matter? Do individual communication rules
like being balancing or being curious matter? To focus our analysis we
concentrate on possibilities and strategies to foster or prevent
consensus.

The above-mentioned models on opinion dynamics, i.e. WD model and HK model, 
were previously studied with the more 
general technique of differential equations on density based state 
spaces instead of single agents in finite populations\cite{Ben-Naim2003,Fortunato2005b,Lorenz2006}. 
However, we will apply these questions to populations of finite size, more
precisely, less than one thousand members.  This prevents us from using such general techniques that abstract from single agents.
Nevertheless, it is interesting because it is a more realistic assumption. The 
model assumes a completely mixed population where everybody has the same
chance of interacting with everybody else. 
However, in human societies the size of groups, where one can reasonably assume a
complete mixing, does not scale arbitrarily. For instance,
Zhou et al. \cite{ZhouEtal2005} and Hill and Dunbar \cite{HillDunbar2003} argue that 
some group sizes are more
frequently observed than others and that at certain critical
number, groups exhibit significantly different properties in,
for instance, their communication patterns. 
The organizational literature also suggests that beyond critical 
sizes, hierarchies will be established. Furthermore, the distribution of people across different geographical locations also restricts the set of potential interaction partners. 
All these aspects suggest that for very large systems the assumption of completely mixed societies is strongly violated. We believe that 
these arguments demonstrate the necessity of investigating 
finitely sized groups if one sticks to the complete mixing assumption.\footnote{
One should be aware that 
research regarding certain social and economic phenomena are only
based on the assumption of finite sizes, e.g. theory on 
competition among firms. In fact, the infinite size assumption 
may sometimes represent the most uninteresting case. As such we suggests that 
in the social sciences the infinite size assumption 
should not be treated as the undiscussed default. At the very least, a justification
that violating the finite size assumption does not cause 
a major change in system behavior is critical.} 
Although finite size is often associated with a difficult analysis, we will demonstrate that for opinion dynamics this approach in some circumstances still lends itself to an analytical approach. 

After a short introduction of the Weisbuch and Deffuant model, 
we answer the question concerning the
extent to which the communication regime is able to enforce or prevent
consensus. We will see that the results
of Deffuant, Weisbuch, and others are not robust against
manipulation of the communication order. The result on
preventing consensus supports Fortunato's claim of universality 
of the threshold for complete consensus \cite{Fortunato2004a}. 
Fortunato provides simulation-based evidence that consensus 
is reached for $\eps>0.5$ irrespective of the structure of an 
underlying connected social network. Our result will explicitly
define a threshold such that for larger bounds of confidence
consensus cannot be prevented. In this way the simulation results
by Fortunato are formally proved without any simulation, but in the
limit of large numbers of agents and uniformly distributed initial
opinions. However, we also show how the result differs for populations
of different finite sizes. For instance, for $\eps$ below a specific
level there is a zero probability of consensus in finite populations, while the level depends on the groups size and the cautiousness of agents.

The communication regime we construct in section \ref{sec:math}
to reach the extreme bounds for preventing and enforcing consensus
relies on full knowledge of the opinions of all agents. To circumvent this, 
in section \ref{sec:sim} we run a simulation analysis with
individual strategies, where agents are either balancing or curious. 
This only requires that agents know their individual recent communication history.
'Balancing' means that an agent who has talked with somebody who
has a higher opinion seeks later on somebody with a lower one.
'Curious' means that agents seek partners with opinion in the same direction as
those of their former communication partners. Particularly the interplay 
of these strategies with the cautiousness that agents exhibit is interesting. 
We will see that these very
simple communication strategies, that could reasonably be applied
by humans, can significantly increase the chances for consensus.

\section{Dynamics of continuous opinions} \label{sec:wd}

We analyze the model of continuous opinion dynamics that was
introduced by Weisbuch, Deffuant, and others \cite{Deffuant2000,
Weisbuch2002}. The dynamics are driven by random encounters of two
agents, who compromise if their distance in opinions is below a certain
bound of confidence $\eps$. The model always converges to a
stabilized opinion formation, where agents in the same cluster
have the same opinion in the long run \cite{Lorenz2005}.

We consider $n\in\N$ agents, who each have an opinion that is represented by a real
number. The opinion of agent $i\in\n:=\{1,\dots,n\}$ at time step
$t\in\N_0$ is represented by $x_i(t) \in \R$. We call the vector
$x(t) \in \R^n$ the \emph{opinion profile} at time step $t$.

\begin{definition}[WD model]
Given an initial opinion profile $x(0) \in \R^n$, a bound of
confidence $\eps\in\R_{>0}$, and a cautiousness
parameter\footnote{This parameter is called convergence parameter
in \cite{Deffuant2000}.} $\mu \in ]0,0.5]$ we define the
\emph{WD model} as a process of opinion dynamics as the random
process $(x(t))_{t\in\N_0}$ that chooses in each time step
$t\in\N_0$ two agents $i$ and $j$ randomly and equally distributed from the set of agents $\n$. 
Agents $i$ and $j$ perform the action
\begin{eqnarray*}\label{def:wd}
&&\hbox{if $|x_i(t)-x_j(t)|<\eps$} \\
&&\quad\quad x_i(t+1) = (1-\mu)x_i(t)+\mu x_j(t), \\
&&\quad\quad x_j(t+1) = \mu x_i(t)+(1-\mu) x_j(t), \\
&&\hbox{else}\\
&&\quad\quad x_i(t+1) =x_i(t), x_j(t+1)=x_j(t).  \\
\end{eqnarray*}
\end{definition}

The bound of confidence $\eps$ was previously shown to be the most
significant parameter to control the number of emerging clusters.
For randomly distributed initial profiles with opinions between
zero and one $x\in [0,1]^n$ and $n = 1000$ it is shown via simulations
that consensus is reached nearly in every case for $\eps
> 0.3$ \cite{Deffuant2000}. For lower $\eps$ the usual outcome is
polarization into a certain number of opinion clusters. Weisbuch,
Deffuant et al.\ derived by computer simulation the
'$1/2\eps$-rule', which states that the number of surviving
clusters is roughly the integer part of $1/2\eps$.\footnote{Very
small surviving clusters are neglected by this rule, but their
existence is systematic as shown by the analysis of a rate
equation for the density of opinions \cite{Ben-Naim2003}.}

The cautiousness parameter $\mu$ had been considered to have no
effect on clustering in the basic model (only on convergence time)
\cite{Deffuant2000,Weisbuch2002}. However, there is already some
evidence that $\mu$ can affect the clustering as well as that the
effect of $\mu$ interacts with other parameters, e.g. number of
agents that participate in an interaction \cite{Urbig2004,Deffuant2006}.
Furthermore, different random initial profiles may lead to
different numbers of clusters, and even the same initial profile
may lead to different numbers of clusters for different random
choices of communicating pairs. In most previous studies the
dependence on the initial profile and on the communication regime is
not considered due to the randomness assumption. In the next section we
will incorporate both, initial opinion profile and communication regime, 
to examine the bounds for enforcing and preventing consensus.

\section{Enforcing and Preventing Consensus} \label{sec:math}

In this section we give mathematical answers to the questions: How
small may $\eps$ be such that enforcing consensus is still possible?
How large may $\eps$ be such that preventing consensus is still
possible?

Let our initial opinion profile $x(0)$ and the parameter
$\mu$ be fixed. We define $\epslow$ as the smallest value of epsilon
for which there is a communication regime that leads to a
consensus. Obviously, $\epslow$ depends on the initial opinion
profile and perhaps on $\mu$. We will give a lower and an upper limit for
$\epslow$ based on a communication regime that looks like a phone
chain of those persons with the most similar opinions, or in other 
words a {\em phone chain of closest}. 

For our approximation we must take a detailed
look at the initial opinion profile. For this reason we regard our 
initial opinion profile $x(0)$
as ordered such that $x_1(0) \leq \dots \leq x_n(0)$, without
loss of generality.
For our considerations it is useful to look at the gaps between
the opinions. We define for $i \in \underline{n-1}$ the gap to the
next neighbor as $\Delta x_i(t) := x_{i+1}(t) - x_{i}(t)$. If we
regard an opinion profile as a function $x_{(\cdot)}(t):\n \to \R$
then we can consider $\Delta x(t) \in \R^{n-1}$ as the discrete
derivative of $x(t)$ with respect to the agent index $i$. $\Delta$ is thus
not a differential but a difference operator. For abbreviation we further
define the \emph{maximal gap} $\max \Delta x := \max_{i\in\underline{n}} \Delta x_i$. 
In our setting with ordered initial opinions, the function
$x_{(\cdot)}(0)$ is monotonously increasing. Thus its difference
function $\Delta x_{(\cdot)}(0)$ is non-negative.

We are now able to define the {\em phone chain of closest} as
a communication regime, which will later on guide us to a fair approximation
of $\epslow$.

\begin{definition}[phone chain of closest]
Let $n\in\N$ be the number of agents. A WD model of opinion dynamics 
is ruled by a \emph{phone chain of closest} 
if the communicating agents at time step
$t\in\N$ are $(t\ \modd\ (n-1))+1$ and $(t\ \modd\ (n-1))+2$.
\end{definition}

The phone chain of closest is
$(1,2),(2,3),(3,4),\dots,(n-1,n),(1,2),$ and so forth. This
sequencing communication strategy provides a nice proof for the
following proposition.

\begin{proposition} 
Let $x(0) \in \R^n$ be an ordered initial
profile and let $\mu \in ]0,0.5]$. It holds that 
\begin{equation}
\max\Delta x(0) \leq \epslow \leq \max_{i\in\underline{n-1}}
\sum_{j=0}^{i-1} \mu^j\Delta x_{i-j}(0).
\end{equation}
\end{proposition}

For a proof see appendix \ref{proof1}. Figure \ref{figChain} shows
how the phone chain of closest works.\footnote{We suspect that 
our estimate is not strict, because we also studied regimes other
than the phone chain of closest. However, the phone chain of closest 
delivers the best result we are able to prove analytically. Yet,
the question of the strictly lowest $\epslow$ is still open.}

\begin{figure}
\centerline{\includegraphics[width=10cm]{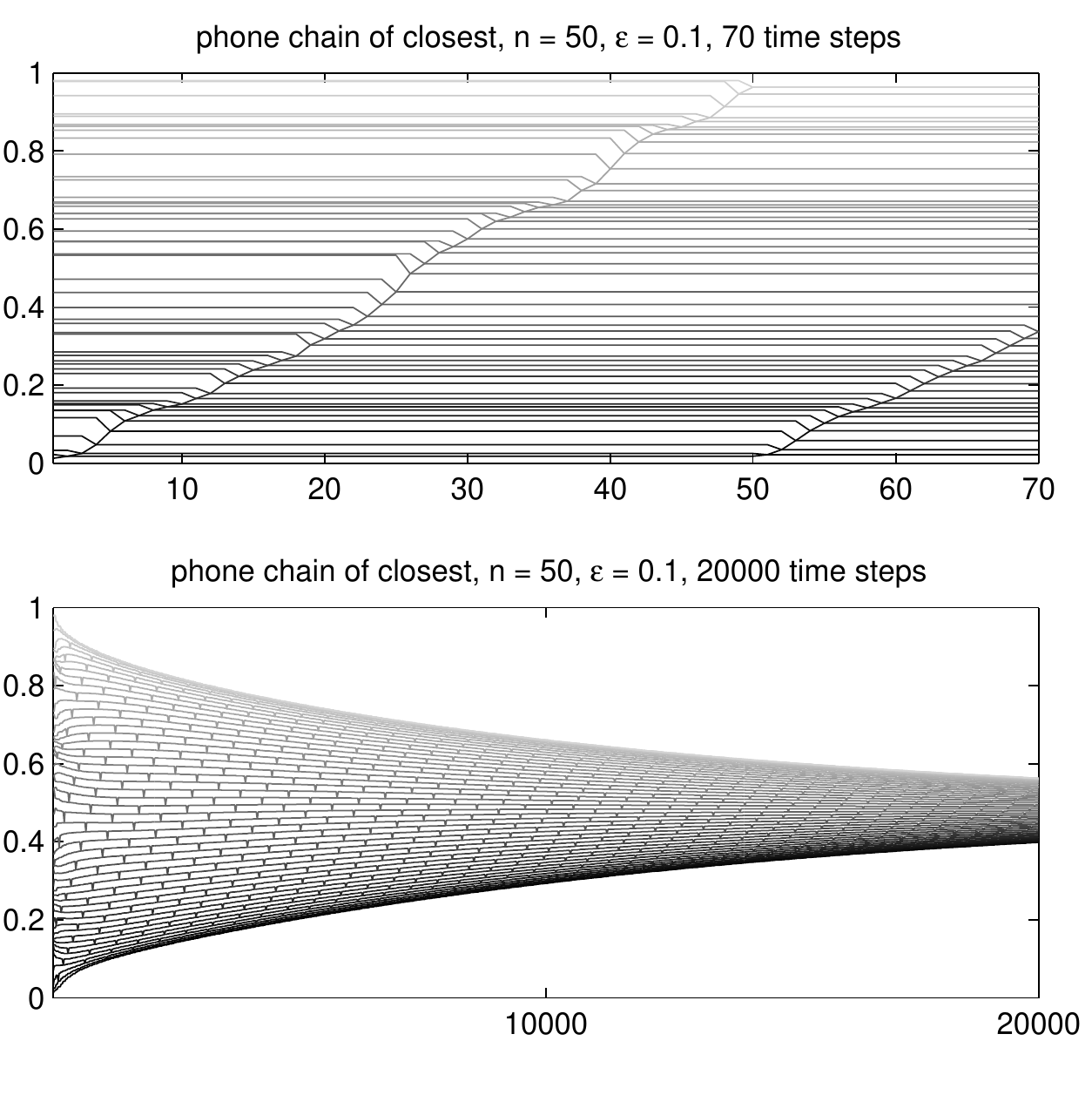}} \vspace*{8pt}
\caption{The phone chain communication regime of closest.}
\label{figChain}
\end{figure}

If we define $\range(x) = \sum_{i\in\underline{n-1}} \Delta x_i =
x_n - x_1$ and $\lceil\cdot\rceil$ as rounding a real value to the upper
integer, then we can derive a corollary with a simpler but not as
sharp bound.

\begin{corollary}
If proposition 1 holds, then it also holds that
\begin{equation}\epslow \leq \frac{1-\mu^{\lceil\frac{\range(x(0))}{\max\Delta x(0)}\rceil}}{1-\mu}\max\Delta x(0)\end{equation}
\end{corollary}

For a proof see appendix \ref{proofcor}.

From corollary 1 one can see that $\epslow$ is determined mostly by the maximal gap, $\mu$, and
the ratio of the maximal gap and the difference between the two
most extreme opinions in the initial profile. For $\mu = 0.5$ the
estimate shows, that enforcing consensus is always possible for
$\eps$ which is twice the maximal gap of the initial profile.

Simulation-based studies often use initial profiles $x(0) \in [0,1]^n$ with
random and uniformly distributed opinions. The length of the
maximal gap in such a profile can be estimated by Whitworth's
formula (\ref{whitworth}) and is thereby dependent on the number
of agents.

\begin{equation}
P(\max\Delta x > \eps) = \sum\limits_{k =
1}^{\lfloor\frac{1}{\eps}\rfloor} (-1)^{k+1} (1-k\eps)^{n-1}
{n\choose k}. \label{whitworth}
\end{equation}

\begin{figure}[htb]
\centerline{\includegraphics[width=10cm]{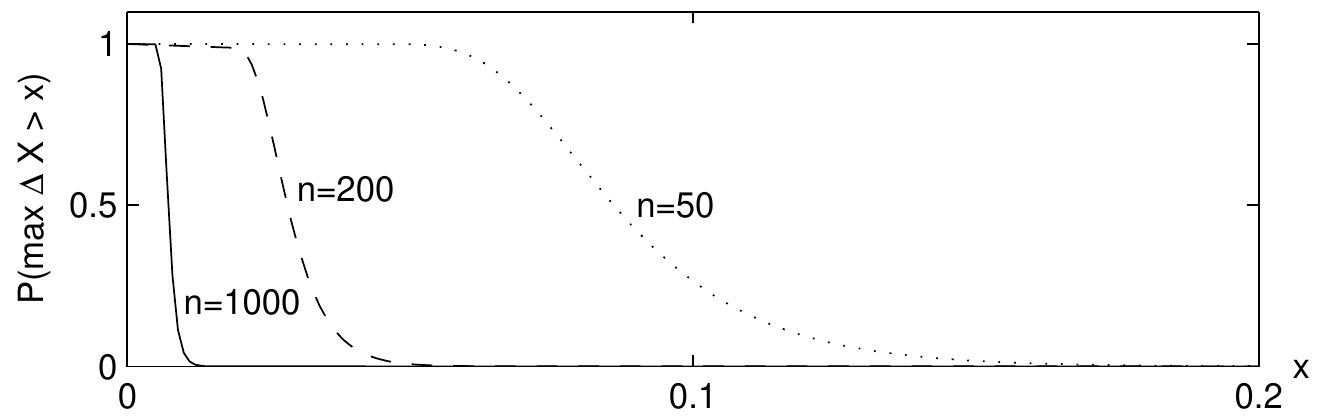}} \vspace*{8pt}
\caption{$P(\max\Delta x > \eps)$ for random equally distributed
$x\in\R^n$ for some $n$.} \label{figgapmax}
\end{figure}

In terms of statistical theory the formula is about the spacings
in an order statistics of $n$ independent uniformly distributed
random variables (see \cite{Devroye1981}). Figure \ref{figgapmax}
shows the probability that the maximal gap is larger than
$\eps$ with $\eps\in[0,1]$.

Based on this distribution it is possible to derive an estimate
for the expected size of the maximal gap in an initial opinion
profile. But it gives an additional insight: the larger the
population the smaller the expected size of the maximal gap. This
leads to the conclusion that for a very large number of agents who
are equally distributed, consensus is possible for extremely low
values of $\eps$. If we assume that the maximal gap converges
to zero as the number of agents increases, then it is possible to reach
consensus for every $\eps$ with a large enough number of agents. However,
a reasoning on infinitely many agents is not appropriate since every 
real society is finite in size. This finite size assumption 
is where many analytic approaches reach their limit.

We now ask the other way around: How large may $\eps$ be such that
preventing consensus is still possible? At first we see that this
question is not detailed enough to be interesting. Preventing
consensus is obviously possible if we forbid one agent to
communicate with all others, e.g. by underlying a disconnected
social network. The right question is: How high may $\eps$ be such
that preventing consensus is still possible, even if we switch at
some time step to an arbitrary communication regime? The biggest
possible $\eps$ is called $\epshigh$.

\begin{proposition} 
Let $x(0)\in\R^n$ be an ordered initial profile and let
$0<\mu<0.5$.
\begin{equation} \label{prop2_1}
\epshigh =
\max_{k\in\underline{n-1}}\left(\frac{1}{n-k}\sum_{i=k+1}^n x_i(0)
- \frac{1}{k}\sum_{j=1}^k x_j(0)\right)
\end{equation}
\end{proposition}

For a proof see appendix \ref{proof2} and lemma 1, which states that
the mean opinion is conserved by the process of opinion formation.

If we regard random and uniformly distributed $x_i(0) \in [0,1]$
for an $n$ approaching infinity, then $\epshigh$ is computed as the
distance of the central points of two arbitrary disjoint intervals
whose union is $[0,1]$. Thus, $\epshigh \to 0.5$ as
$n\to\infty$. This proves the Fortunato's universality result
\cite{Fortunato2004a} by showing that
preventing consensus is impossible for a large enough number of
connected agents for $\eps > 0.5$. Furthermore, Fortunato delivers evidence that consensus is not possible for $\eps<0.5$  as $n\to\infty$ and random pairwise communication regardless of an underlying network topology. Proposition 1 shows that there are specific communication orders that lead the society to consensus even for very low values of $\eps$ for every finite but arbitrarily large number of agents. However, the probability of obtaining one of these consensus-enforcing communication orders when picking it out of the set of random pairwise communication orders would probably approach zero in the limit of large $n$. Hence, if we are free to choose or to influence the communication order, then Fortunato's claim that consensus is not possible for $\eps<0.5$ is disproved. It remains to prove the impossibility of consensus for $\eps<0.5$ under random pairwise communication in the sense that our consensus enforcing communication orders approach a probability of zero as $n$ increases. 

\begin{figure}[htb]
\centerline{\includegraphics[width=10cm]{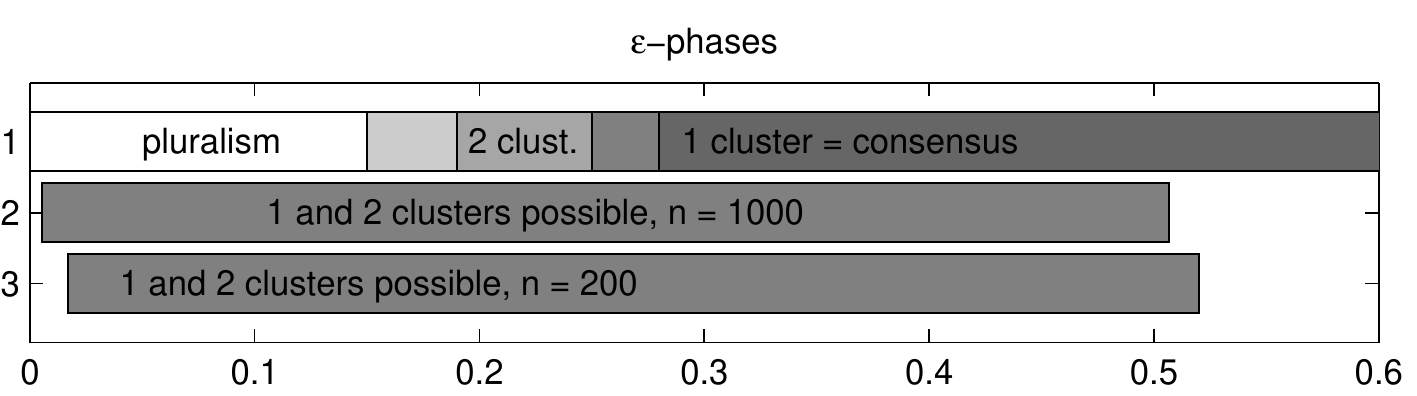}}
\vspace*{8pt} \caption{Numerical $[\epslow,\epshigh]$ (second
darkest gray) where consensus and polarization occurs (1) and is
possible (2 and 3).} \label{possibilityRange}
\end{figure}

The interval $[\epslow,\epshigh]$ is the range where both
enforcing and preventing consensus is possible with an appropriate
communication regime. Both bounds of the interval depend on the
initial profile $x(0)$. Figure \ref{possibilityRange} provides some
numerical evidence about the possibilities that can be reached
with manipulation of the communication regime. The data in line one
come from figure 4 in \cite{Deffuant2000} (visually extracted)
with 250 simulation runs with random and equally distributed
initial profiles and $n=1000$. For line two we took 250 randomly chosen and 
equally distributed profiles with $n=1000$ and show the maximal
$\epslow$ and minimal $\epshigh$ that occurred in all 250 profiles
(all computed with propositions 1 and 2). Enforcing and
preventing consensus was possible in the displayed interval for
all 250 selected profiles. The same is true in line three for $n=200$.

From this figure we can also see that for finite populations and random
initial profiles avoiding consensus is possible even for $\eps >
0.5$ (in contrast to Fortunato). The larger the population the
smaller $\eps$ can be while still guaranteeing the possibility of
consensus. The larger the population the smaller is the upper
limit of $\eps$ for which consensus can be prevented.

%
%
%

\section{Individual strategies that increase chances for
consensus}\label{sec:sim}

In the previous section we applied a mechanism for enforcing consensus that was built on knowledge about all people's opinions. We now want to leave behind this idea of global knowledge and the great
master plan for communication and go to agent based strategies, which 
may also promote consensus. Our agents do
not know the opinions of all other agents and thus do not know if they
are in the center or at the extremes of the opinion space. The agents
follow rules that only require knowledge of their own communication history.

\subsection{Balancing and curious agents with directions}

From the huge set of possible individual communication strategies 
we focus on balancing and curious agents. Consider an agent who has communicated with another agent and adapted his opinion accordingly. A {\em balancing agent} will now search for an agent whose opinion is contrary to that of the previous communication partner, while ignoring all other agents. A \emph{curious agent} will instead seek out a new communication partner whose opinion is in line with that of the previous communication partner, again ignoring other agents.

To prevent agents from not finding an agent to compromise with,
we introduce a new parameter, $f_{\max}$, which represents a maximal level of frustration. 
Specifically, it is the number of unsuccessful attempts
a agent sticks to the rule before abandoning it. Thus, agents
are not forced to follow the strategies for ever. We store relative opinions of potential communication partners, more precisely, the direction, and individual frustration levels for all agents in a vector $d\in\mathbb{Z}^n$. If $d_i$ is negative, then agent $i$ wants to
compromise with an agent with a lower opinion. If $d_i$ is positive, then 
agent $i$ wants to compromise with an agent with a higher opinion.
If $d_i$ is zero agent $i$ has no preferred direction. The
absolute values of the directions represent the frustration level. The lower the absolute value the higher is the frustration; if it reaches zero, then the agents does not care anymore about the direction of potential communication partners.

Frustration and direction are additional factors that affect agents tendency to compromise.
Agents $i$ and $j$ only compromise if both agents' opinions are in the corresponding set of opinions the other agent looks for, i.e. $d_i*d_j \leq 0$. If they are not in the set, but are closer than
$\epsilon$ to each other, then they both reduce the absolute value of
their frustration levels each by one point. Thus the absolute values of $d_i$ and $d_j$ decrease. After a
successful compromise, agents set $d_i$ to
$f_{\max}$ with the sign indicating the new search direction.
Curious agents differ from balancing agents in the sign of $d_i$. 
Besides this restriction we return to random pairwise communication.
The corresponding pseudo-code can be found in \ref{subsec:dir}.

Societies of balancing and curious agents are essentially identical in
their dynamics if $\mu=0.5$. It is interesting to note that after a compromise between
two balancing agents we end up with two agents with the same opinion
searching in opposite directions; however, the same applies to
curious agents but with agents whose indices are reversed. Thus,
clustering outcomes are identical for both balancing and curious agents when $\mu=0.5$.

\subsection{Simulation setup}
For both strategies, balancing and curious agents, we ran
simulations for the values $\mu = 0.2, 0.5$, $n=50,100,200$, 
$\eps=0,\stackrel{+0.01}{\dots},0.35$, and $f_{\max} =
0,1,2,4,8,16,32$. For each point in this parameter space we have
3000 independent simulation runs with random initial profiles and
random selection of communication partners.\footnote{To check
larger numbers of agents, we performed 3000 independent simulation runs
for balancing agents for $n=500,1000$,
$\eps=0,\stackrel{+0.01}{\dots},0.35$, $\mu=0.2,0.5$, and $f_{\max}
= 0,1,2,4,8$.}

Each simulation run stops when we reach a configuration where 
all indirectly connected\footnote{Two
agents are connected if their opinions differ by not more than
$\eps$. They are indirectly connected if there is a chain of
connected agents between them.} subgroups of agents have
a maximal opinion difference smaller than $\eps$ and thus cannot further
split. 
The mean preserving property (see
lemma 1 in \ref{proof2}) of the dynamics permits a calculation of the
long term limit of the convergence process. 
We consider the
\emph{average size of the biggest cluster} after stabilization as
a measure for the possibility of consensus.\footnote{Another
possible measure would be the average number of clusters. But the
Weisbuch and Deffuant model is known to produce minor clusters of only a few agents
\cite{Ben-Naim2003,Lorenz2005a}.} All simulations were implemented using ANSI-C. 
The program code is available on request from the second author.

\subsection{Simulation results}

\begin{figure}
\centerline{\includegraphics[width=12cm]{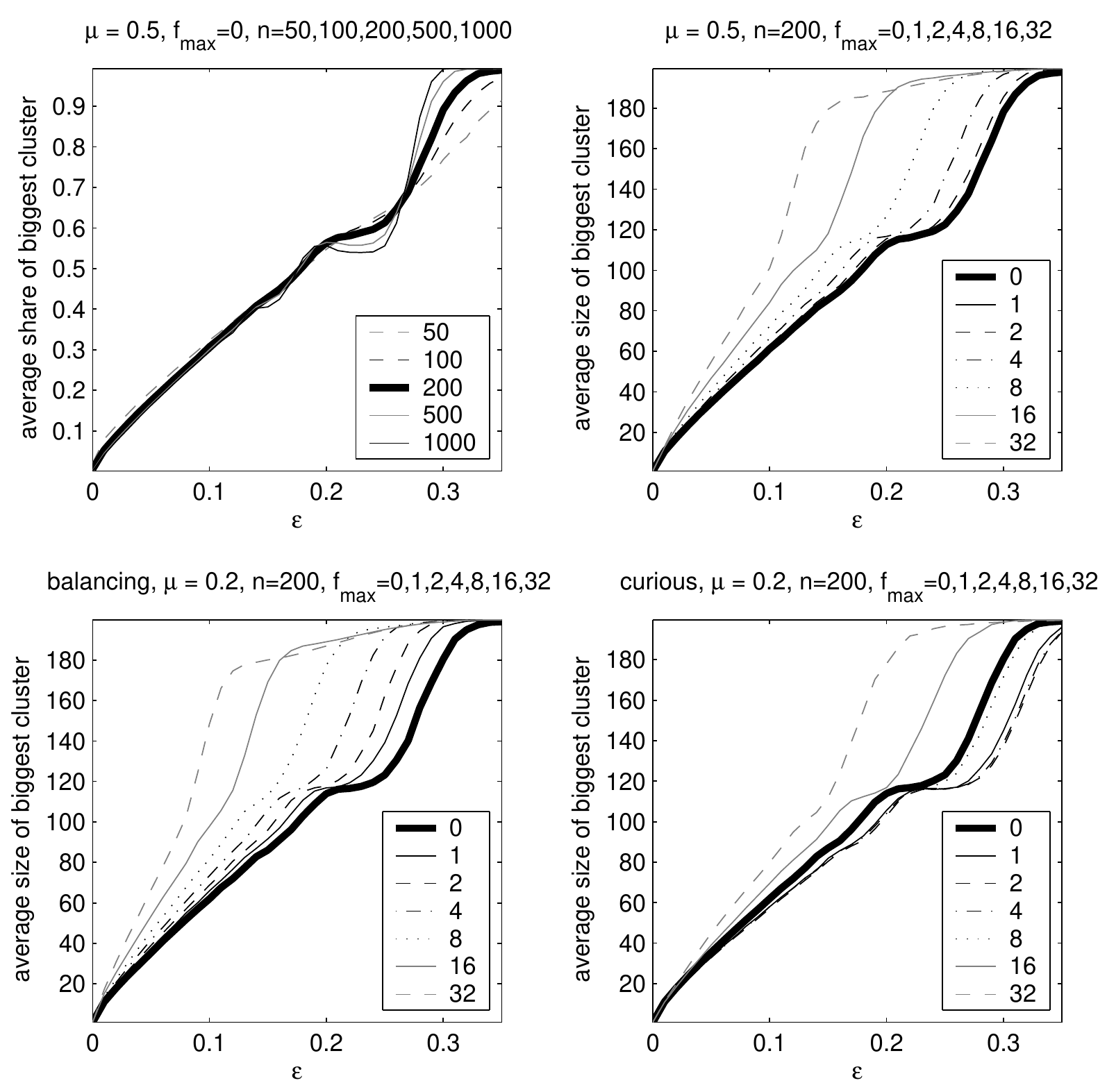}} \vspace*{8pt}
\caption{The average size of the biggest cluster for initial
profiles with 50, 100, 200, 500 and 1000 agents (top left) and
$\mu = 0.5$, for 200 balancing/curious agents with $\mu = 0.5$
(top right), for 200 balancing agents with $\mu = 0.2$ (bottom
left), and 200 curious agents with $\mu = 0.2$ (bottom right).
} \label{figChainDV}
\end{figure}

\begin{figure}
\centerline{\includegraphics[width=12cm]{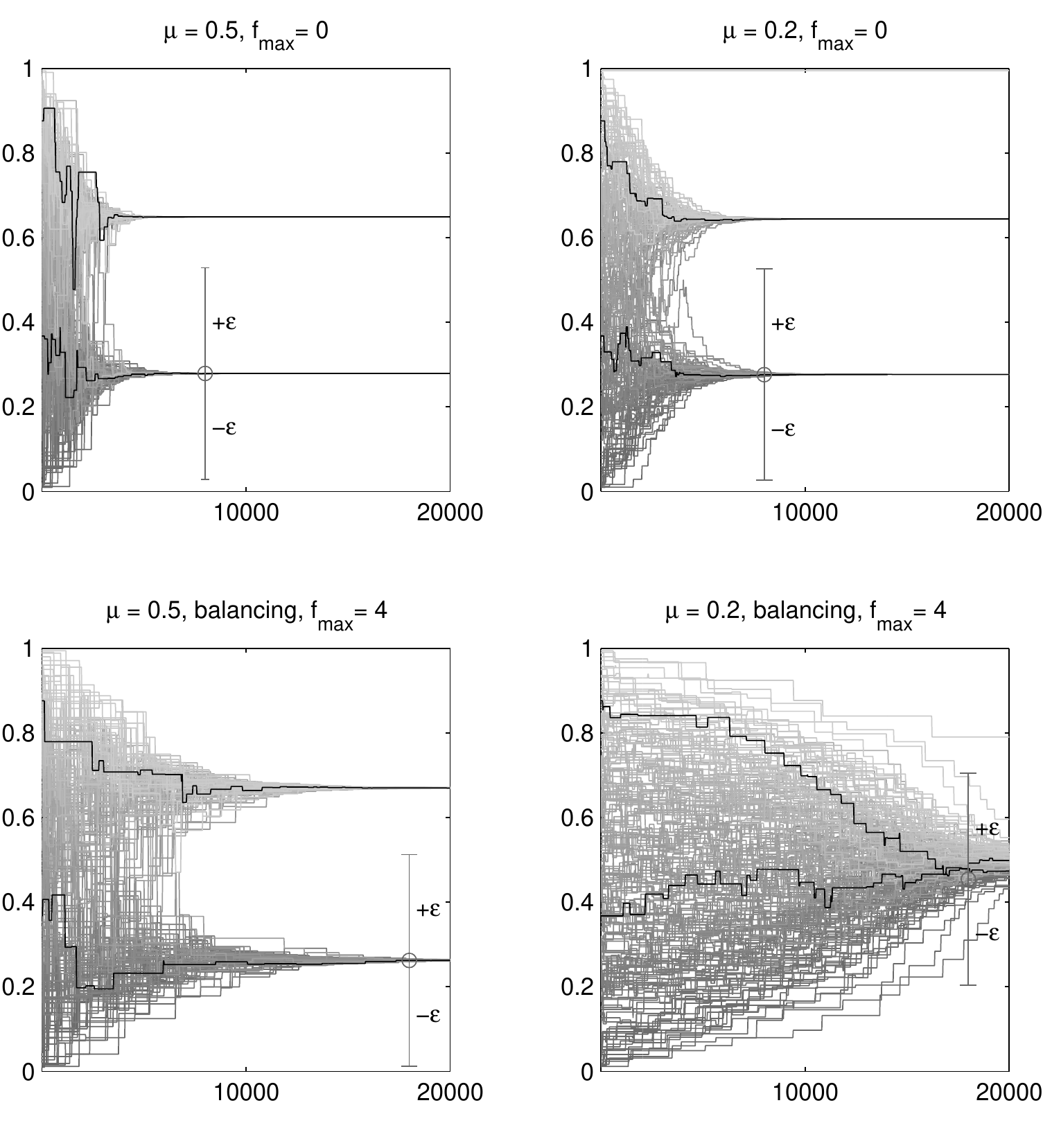}} \vspace*{8pt}
\caption{Example processes for balancing agents with
$n=200$ and $\eps=0.25$. This demonstrates that cautiousness fosters the
positive effect of balancing.} \label{figExBal}
\end{figure}

\begin{figure}
\centerline{\includegraphics[width=12cm]{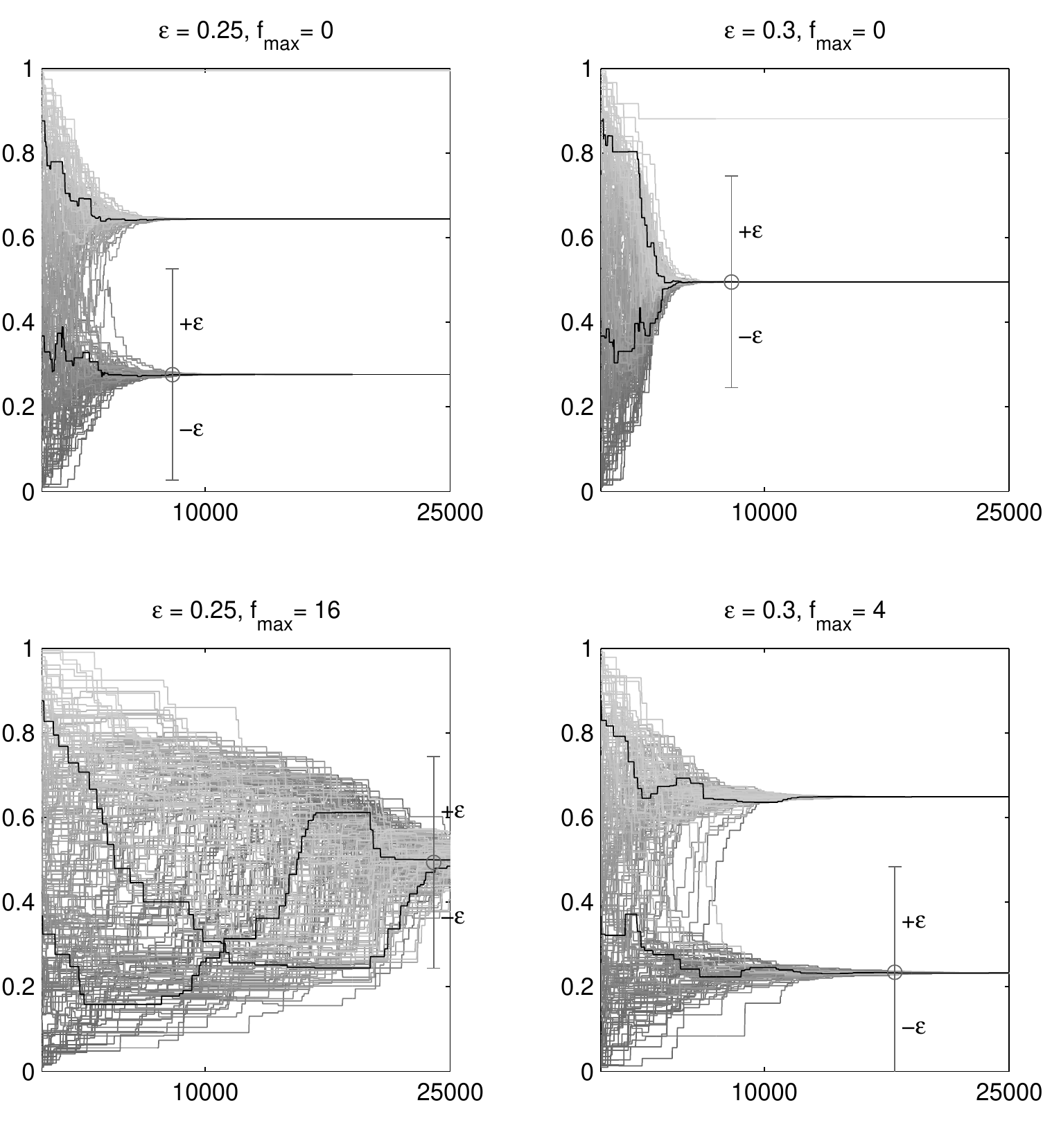}} \vspace*{8pt}
\caption{Example processes for curious agents with
$n=200$ and $\mu=0.2$. This demonstrates that for cautious agents being
curious with a high frustration level one can foster consensus (left
hand side) but for a low frustration level it may destroy consensus
(right hand side).} \label{figExCur}
\end{figure}

Figure \ref{figChainDV} shows the results for the average size of
the biggest cluster. The fat line always represents the average size of
the biggest cluster for $f_{\max}=0$ and $n=200$. The upper left plot
shows how this line changes for varying numbers of agent. The
plateau at $\eps=0.2$ shows the characteristic polarization phase
where agents form two big clusters (see for instance in
\cite{Deffuant2000}). We see that this plateau becomes more pronounced
 for larger $n$ and less distinct for smaller
$n$.\footnote{The small 'hill' at $\eps = 0.19$ for $n=500, 1000$
is another interesting phenomenon related to the measure of the average size of the
biggest cluster, but it is outside of the scope of this paper.}

Next, we examine how the
transition from polarization to consensus shifts to other
$\eps$-regions when agents are balancing or curious. We
concentrate our discussion on the case of 200 agents, but we checked that the shifts for
$n=50,100,500,1000$ are similar.\footnote{Data for $n=50,100,500,1000$ is available on
request from the second author.}

The upper right plot in figure \ref{figChainDV} shows the effect of
balancing and curious agents for $\mu =0.5$, in which case these two types exhibit the same dynamics. 
In the lower
plots we distinguish between balancing and curious for agents who are more
cautious, i.e. $\mu = 0.2$. The thin lines show the effects
of an increase in the maximal frustration $f_{\max}$ under a given strategy and $\mu$. 
The main conclusions from figure \ref{figChainDV} are: Being balancing
has a positive effect on the chances for consensus. For
$\mu=0.5$ this holds for all maximal frustrations $f_{\max} > 1$.
The same holds trivially for curious agents under $\mu = 0.5$. A
smaller $\mu$, which means being more cautious, supports the positive
effects for balancing agents. However, a smaller $\mu$ does not support
the positive effects of curious agents.\footnote{An interesting
but small effect is that the general tendency 
of an increase in chances of consensus with an increase in 
$f_{\max}$ is sometimes slightly violated. For instance, for balancing agents with $\mu=0.2$ and
$f_{\max}=16$ we observe a slightly larger average size of the
biggest cluster than for $f_{\max}=32$. We suspect that this is a systematic effect and not caused by chance. However, the effect is so small that we did not further study the causes.
 }

Figure \ref{figChainDV} is based on aggregated data. To give a more detailed picture of the dynamics, figures \ref{figExBal} and \ref{figExCur} show some single simulation runs.  
Figure \ref{figExBal} shows how balancing agents with an
intermediate frustration maximum are positively effected in
finding a consensus by being more cautious.
Figure \ref{figExCur} shows the ambivalent effects of curious
agents who are cautious. While a high frustration maximum can
foster consensus, an intermediate frustration maximum may even
prevent consensus. We see that almost every curious agent has to
cross the central opinion if curious agents want to reach a
consens.
Figures \ref{figExBal} and \ref{figExCur} also illustrate that more chances for consensus
by being balancing or curious is always paid by a
longer convergence time.

\section{Discussion}

Our analytical results describe the possibilities of consensus 
in the Weisbuch and Deffuant model and we prove the universality 
of the consensus threshold in the sense of
Fortunato \cite{Fortunato2004a}. Both enforcing and preventing
consensus is possible in a large interval for values of $\eps$ and we give an
impression of how it scales with the number of agents and the
cautiousness parameter. This shows the large impact that the 
control of communication has on consensus
formation in the Weisbuch and Deffuant model in finite populations.
Communication control is a feature of real opinion dynamics, which
is to some extent manipulable through organizations. Therefore, our 
results are of interest for those who aim at designing communication and
discussion processes and want to foster consensus or dissent (see for example \cite{Lorenz2005c}).

Generally, continuous opinion dynamics under bounded confidence is
driven by the opposition of the consensus-promoting force of averaging and
the separating force of bounded confidence. 
Dynamics start at the extremes of the opinion space. Specifically, the most extreme agents move towards less extreme position and thus higher densities of opinions evolves at both
extremes. These two high density regions attract agents from the
center and may lead to a split in the opinion range.

We explored by simulation the dynamics of societies where each
individual behaves balancing or curious. Balancing agents tend to
move in a narrow zigzag around their first opinion, while curious
agents tend to move in a wide zigzag exploring almost the whole opinion
space. Therefore, both strategies have a tendency to prevent a rapid clustering. 
Balancing agents do this by seeking input from both sides, 
which prevents them from quickly being absorbed by a nearby cluster.
Curious agents tend to run through and finally break out of
a cluster they recently joined. They tend to explore more of the opinion
space. 
Since only the clustering is slowed down, while the overall contraction process of the opinion profile 
keeps its speed, the chances for consensus are increased.

We further extend the analyses in \cite{Urbig2004,Deffuant2006} 
on the role of the cautiousness parameter $\mu$.
Particularly in the first part of this paper we see that 
cautiousness significantly controls the possibility of consensus. 
Furthermore, the second part of the paper illustrates the intriguing interplay
of this parameter with agents' communication strategies.  

To summarize, if you
want your agents to foster consensus by balancing, you should
appeal to them to be cautious. If you want them to foster
consensus by being curious you should appeal to them not to be
cautious, otherwise you may even get a negative effect when agents
have a low frustration maximum.
This results from the fact that balancing agents prevent clustering
by trying to avoid early absorption into a clusters and thus smaller
steps, i.e. smaller $\mu$, have a positive effect on formation of consensus. 
Curious agents prevent clustering
by getting out of clusters they recently entered; hence, smaller steps
have a negative effect. In general, the impact of being balancing is higher than that of
being curious, yet both individual strategies can foster
consensus.

\begin{appendix}

\section{Appendix for proofs}

\subsection{Proof of Proposition 1} \label{proof1}
{\bf Proof. }
The left inequality results from the fact that an
$\eps<\max_{i\in\underline{n-1}}\Delta x_i(0)$ can obviously not
bridge the maximal gap; thus, the opinion profile will be divided
into the two groups above and below this gap forever, regardless
of any communication structure.

To show the right inequality, let $\eps > \max_{i\in\underline{n-1}}
\sum_{j=0}^{i-1} \mu^{j}\Delta x_{i-j}(0)$. We will show that our specific 
communication regime, the
phone chain of closest, drives the dynamic to a
consensus.

First, the phone chain of closest can not change the order of the
opinion profile. Thus, it holds for all $t\in\N_0$ that $x_1(t) \leq \dots \leq x_n(t)$.

In a first step we will look at the $n-1$ first time steps, thus 
the first phone chain round. After one round we will see that
the maximal gap in $\Delta x(n-1)$ has shrunk substantially and we
can conclude with an inductive argument.

Let us consider that there is no bounded confidence restriction by
$\eps$, thus in every time step two opinions really change (if
they are not already equal). We will derive equations for $\Delta
x$ in the time steps $1,\dots,n-1$ under this assumption. After
that we will see that $\eps$ does not restrict this dynamic.

Let $i\in\underline{n-1}$ be an arbitrary agent. We focus on
$\Delta x_i$ the gap between $i$ and $i+1$ for all time steps and
will deduce formulae only containing values of the initial
profile. Agent $i$ at time step $i-1$ has communicated recently
with agent $i-1$ and will communicate with agent $i+1$. Thus
\begin{equation}
\Delta x_i(i-2) = \dots = \Delta x_i(1) = \Delta x_i(0)
\label{prop1_1}.
\end{equation}
Due to the communication with agent $i-1$, agent $i$ moves towards
$i-1$ thus $\Delta x_i$ gets larger.
\begin{equation}
\Delta x_i(i-1) = \Delta x_i(i-2) + \mu \Delta x_{i-1} (i-2)
\label{prop1_2}
\end{equation}
By recursion of (\ref{prop1_1}) and (\ref{prop1_2}) it follows that
\begin{eqnarray}
\Delta x_i(i-1) & = & \Delta x_i(0) + \mu\Delta x_{i-1}(0) + \dots
\nonumber \\
& & \dots + \mu^{i-2}\Delta x_2(0) + \mu^{i-1}\Delta
x_1(0) \nonumber \\
 & = & \sum_{j=0}^{i-1} \mu^j \Delta x_{i-j}(0) \label{prop1_3}
\end{eqnarray}
Going one step further to the communication of $i$ and $i+1$ where
their opinion gets closer we get the following:
\begin{equation}
\label{prop1_4} \Delta x_i(i)= \Delta x_i(i-1) - 2\mu\Delta
x_i(i-1)
\end{equation}
We use $\Delta x_{i}(i-1)$ as an abbreviation for the right hand
side of (\ref{prop1_3}) which only contains expressions at time
step 0. 

In the next step $\Delta x_i$ becomes larger as agent $i+1$ moves towards agent $i+2$.
\begin{equation}
\begin{array}{rcl} \Delta x_i(i+1) & = & \Delta x_i(i) +
\mu\Delta
x_{i+1}(i) \\
\stackrel{(\ref{prop1_4})(\ref{prop1_2})}{=}&
\multicolumn{2}{l}{(1-2\mu)\Delta x_i(i-1) + \dots}  \\
    & \multicolumn{2}{l}{\ \ \ + \mu(\Delta x_{i+1}(i-1) + \mu \Delta x_i(i-1))} \\
    \stackrel{(\ref{prop1_1})}{=}& \multicolumn{2}{c}{\mu\Delta x_{i+1}(0) +
    (1-2\mu+\mu^2)\Delta x_i(i-1)} \label{prop1_5}
\end{array}
\end{equation}
To complete all time steps until $t=n-1$ we have to mention
\begin{equation}\label{prop1_6}
    \Delta x_i(i+1) = \Delta x_i(i+2) = \dots = \Delta x_i(n-1).
\end{equation}
For $\Delta x_{n-1}$ there is no equation (\ref{prop1_5}) the
last value after the phone chain round is computed by equation
(\ref{prop1_4}).

To make all these equations valid and thus to ensure that no opinion
change is prevented by $\eps$, it must hold for all
$i\in\underline{n-1}$ that $\Delta X_i(i-1) < \eps$. Looking at
(\ref{prop1_3}) we see that this is the case by construction
of the lower bound of $\eps$.

From Equation (\ref{prop1_3}), (\ref{prop1_5}) and (\ref{prop1_6})
we get
\begin{equation}
\begin{array}{rcl}
    \Delta x_i(n-1) &=& \mu\Delta x_{i+1}(0) + \\
    & & +(1-2\mu+\mu^2)\sum_{j=0}^{i-1} \mu^j \Delta x_{i-j}(0) \\
    \multicolumn{3}{c}{\leq \left(\mu + (1-2\mu+\mu^2)\sum_{j=0}^{i-1} \mu^j\right) \max\Delta x(0)} \\
    \multicolumn{3}{c}{ = \left(\mu +
    (1-\mu)^2 \frac{(1-\mu^i)}{1-\mu}\right) \max\Delta x(0) }\\
    \multicolumn{3}{c}{= (1-\mu^i+\mu^{i+1}) \max\Delta x(0) }\\
\end{array}\label{prop1_7}
\end{equation}

Thus it holds $\max\Delta x(n-1) \leq
(1-\mu^i+\mu^{i+1})\max\Delta x(0)$. 
It is easy to see that $k:=1-\mu^i+\mu^{i+1} < 1$ for $0<\mu<1$.

For the next phone chain rounds we can conclude with the same
procedure and it will hold that $\max\Delta x(t(n-1))\leq k^t
\max\Delta x(0)$. Thus $\max\Delta x(t)$ converges to zero, which
implies that the process converges to a consensus.
\hfill$\Box$

\subsection{Proof of Corollary 1} \label{proofcor}
{\bf Proof. }
With abbreviation $x:=x(0)$ we use the equations $\range(x) =
\sum_{i=1}^{n-1} \Delta x_i$ to derive $\range(x) \leq
\lceil\frac{\range(x)}{\max\Delta(x)}\rceil\max\Delta x$. This
gives

\begin{equation}\label{ee1}\sum_{i=1}^{n-1} \Delta x_i \leq \sum_{i=1}^{\lceil\frac{\range(x)}{\max\Delta(x)}\rceil} \max\Delta(x)\end{equation} and therefore
\begin{equation}\epslow \leq
\max_{i\in\underline{n-1}} \sum_{j=0}^{i-1} \mu^j\Delta x_{i-j}
\leq \sum_{i=0}^{\lceil\frac{\range{x}}{\max\Delta(x)}\rceil}
\mu^{i-1} \max\Delta(x) \end{equation}

Taking the right-hand side of (\ref{ee1}) and transforming it to
$\frac{1-\mu^{\lceil\frac{\range(x)}{\max\Delta
x}\rceil}}{1-\mu}\max\Delta x$ finally provides us with corollary 1. 
\hfill$\Box$

\subsection{Proof of Proposition 2} \label{proof2}

\begin{lemma}
Let $x(0)\in\R^n$ be an initial profile and $(x(t))_{t\in\N_0}$ be
a process in a WD model of opinion dynamics with arbitrary $\eps,\mu$. For
every time step $t\in\N_0$ it holds that
\begin{equation}\label{lemma1}
\frac{1}{n}\sum_{i=1}^n x_i(t) = \frac{1}{n}\sum_{i=1}^n x_i(0).
\end{equation}
\end{lemma}

{\bf Proof. }
Obvious by definition \ref{def:wd}.
\hfill$\Box$

{\bf Proof of Proposition 2.}
Let $\eps \leq \epshigh$. Let us divide the set of agents
according to the maximal $k$ in equation (\ref{prop2_1}) into two
subsets $I_1 = \{1,\dots,k\}, I_2= \{k+1,\dots,n\}$. We choose a
communication regime where both subgroups find their respective consensus
$x_1 = \dots = x_k = c_1, x_{k+1} = \dots = x_n = c_2$. This
should be possible, otherwise we are have established a persistent dissence.
Due to lemma 1 and equation \ref{prop2_1} it holds $|c_1-c_2|\geq\eps$ and
communication is not possible between the subgroups any more.
\hfill$\Box$

\section{Appendix for pseudo code}\label{subsec:dir}

For balancing agents we use:
\begin{verbatim}
 1: initialize X[]
 2: initialize D[] = (0,0,...,0)
 3: WHILE not clustered(X) AND changes possible
 4:    choose agent i,j
 5:    IF |X[i]-X[j]| <= epsilon
 6:        IF (X[j]-X[i])*D[i] >= 0 AND (X[j]-X[i])*D[i] >= 0
 7:            X[i]=X[i] - mu*(X[i]-X[j])
 8:            X[j]=X[j] + mu*(X[i]-X[j])
 9:            D[i] = + sign(X[i]-X[j]) * fmax
10:            D[j] = - sign(X[i]-X[j]) * fmax
11:        ELSE
12:            IF D[i]!=0
13:                D[i]= D[i] - sign(D[i])
14:            ENDIF
15:            IF D[j]!=0
16:                D[j]= D[j] - sign(D[j])
17:            ENDIF
18:        ENDIF
19:    ENDIF
20: ENDWHILE
\end{verbatim}
For curious agents we use same code with signs switched in lines
nine and ten.

\end{appendix}

\bibliographystyle{unsrt}

\end{document}